\documentclass[aps,pra,twocolumn,
	amsmath,amssymb,amsfonts,floatfix,
	longbibliography,superscriptaddress,nobibnotes]{revtex4-2}

\usepackage[utf8]{inputenc}
\usepackage[T1]{fontenc} 
\usepackage{graphicx} 
\usepackage{bm} 
\usepackage{ulem}

\usepackage[hypertexnames=false,colorlinks=true,citecolor=blue,
			linkcolor=blue,urlcolor=blue]{hyperref}

\usepackage{xfrac}

\graphicspath{{fig/}{./fig/}{.}}

\begin{document}

\title{Electronic band structure and surface states in Dirac semimetal LaAgSb$_{2}$}

\author{Marcin Rosmus}
\email[e-mail: ]{marcin.rosmus@uj.edu.pl}
\affiliation{\mbox{Marian Smoluchowski Institute of Physics, Jagiellonian University, Prof. S. {\L}ojasiewicza 11, PL-30348 Krak\'{o}w, Poland}}
\affiliation{\mbox{Solaris National Synchrotron Radiation Centre, Jagiellonian University, Czerwone Maki 98, 30-392 Krak\'{o}w, Poland}}

\author{Natalia Olszowska}
\affiliation{\mbox{Solaris National Synchrotron Radiation Centre, Jagiellonian University, Czerwone Maki 98, 30-392 Krak\'{o}w, Poland}}

\author{Zbigniew Bukowski}
\affiliation{\mbox{Insitute of Low Temperature and Structure Research, Polish Academy of Sciences, P.O. Box 1410,50-950 Wroc\l{}aw, Poland}}

\author{Pawe\l{} Starowicz}
\affiliation{\mbox{Marian Smoluchowski Institute of Physics, Jagiellonian University, Prof. S. {\L}ojasiewicza 11, PL-30348 Krak\'{o}w, Poland}}

\author{Przemys\l{}aw Piekarz}
\affiliation{\mbox{Institute of Nuclear Physics, Polish Academy of Sciences, 
W. E. Radzikowskiego 152, PL-31342 Krak\'{o}w, Poland}}

\author{Andrzej Ptok}
\email[e-mail: ]{aptok@mmj.pl}
\affiliation{\mbox{Institute of Nuclear Physics, Polish Academy of Sciences, 
W. E. Radzikowskiego 152, PL-31342 Krak\'{o}w, Poland}}

\date{\today}

\begin{abstract}
LaAgSb$_{2}$ is a Dirac semimetal showing charge density wave (CDW) order.
Previous ARPES results suggest the existence of the Dirac-cone-like structure in the vicinity of the Fermi level along the $\Gamma$--M direction \mbox{[X. Shi {\it et al.}, \href{http://doi.org/10.1103/PhysRevB.93.081105}{Phys. Rev. B {\bf 93}, 081105(R) (2016)}]}. 
This paper is devoted to a complex analysis of the electronic band structure of LaAgSb$_{2}$ by means of angle-resolved photoemission spectroscopy (ARPES) and theoretical calculations within the direct {\it ab initio} method as well as tight binding model formulation.
To investigate the possible surface states we performed the direct DFT slab calculation and the surface Green function calculation for the (001) surface. 
The appearance of the surface states, which depends strongly on surface, points to the conclusion that LaSb termination is realized in the cleaved crystals. 
Moreover, the surface states predicted by our calculations at the $\Gamma$ and $X$ points are found by ARPES.
Nodal lines, which exist along X--R and M--A path due to crystal symmetry, are also observed experimentally. 
The calculations reveal another nodal lines, which originate from vanishing of spin-orbit coupling and are located at X--M--A--R plane at the Brillouin zone boundary.
In addition, we analyze band structure along the $\Gamma$--M path to verify, whether Dirac surface states can be expected. 
Their appearance in this region is not confirmed.
\end{abstract}

\maketitle

\section{Introduction}
\label{sec.intro}

The discovery of topological insulators with a large gap opened a period of intensive studies of this type of topological systems~\cite{hasan.kane.10, qi.zhang.11,sato.ando.17}.
Due to the intrinsic band inversion it is possible to realize the topologically protected conducting surface states with linear dispersion, called Dirac cones~\cite{hsieh.xia.09,zhang.liu.09,xia.qian.09,alpichshev.analytis.10,kuroda.arita.10}.
Such surface states paved the way for studying a novel phase of matter~\cite{armitage.mele.18,schoop.pielnhofer.18,klemenz.lei.19}.

\begin{figure}[!b]
\includegraphics[width=\linewidth]{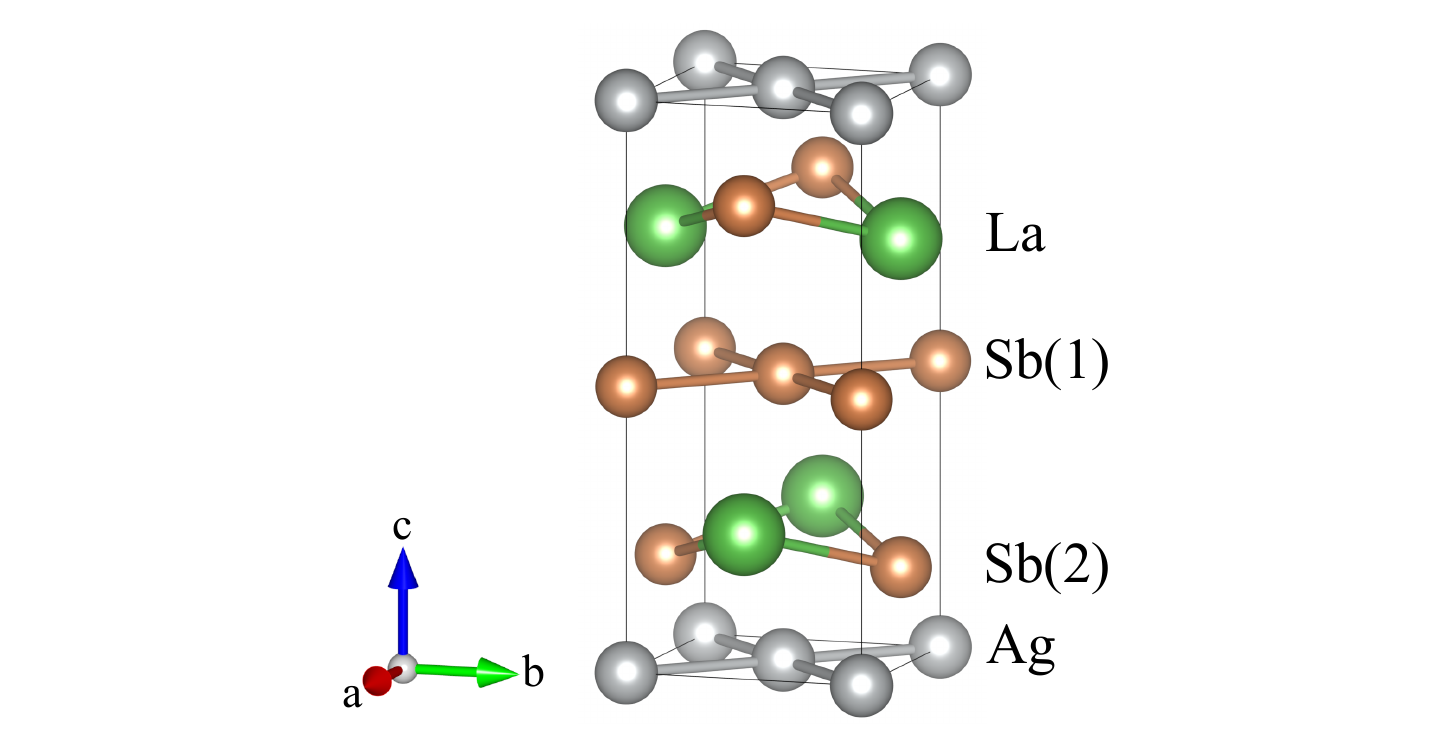}
\caption{
The tetragonal crystal structure of LaAgSb$_{2}$.
}
\label{fig.crys}
\end{figure}

In practice, Dirac cones are observed not only in topological insulators but also in other topological systems.
One of the examples, where the existence of these surface states are expected, is the Dirac semimetal LaAgSb$_{2}$ (Fig.~\ref{fig.crys}).
It is characterized by rare and untypical coexistence of the topological phase and the charge density wave (CDW) order~\cite{song.park.03}.
The CDW modulations in LaAgSb$_{2}$ were found by the x-ray scattering measurements~\cite{song.park.03}.
Also, thermal conductivity and $^{139}$La nuclear magnetic resonance (NMR) indicate the phase transition~\cite{lue.tao.07}.
A periodic charge and lattice modulation with the wave vector ${\bm q}_{1} \sim 0.026 \times ( 2\pi/a )$ develops along the $a$ direction below the temperature $T_\text{CDW}^{1} = 207$~K.
Further decreasing of the temperature results in an additional CDW ordering below $T_\text{CDW}^{2} = 186$~K along the $c$ direction with ${\bm q}_{2} \sim 0.16 \times ( 2\pi/c )$.
The realization of CDW with tiny modulation wave vectors can be associated with the Fermi surface (FS) nesting~\cite{bosak.souliou.21}. 
Application of the external hydrostatic pressure, leads to suppression of $T_\text{CDW}$ and finally to disappearance of the CDW phases (around $2-3$~GPa)~\cite{akiba.nishimora.21,budko.wiener.06,torrikachvili.budko.07}.
Similarly, the CDW phase can be destroyed by the chemical doping~\cite{budko.wiener.06,torrikachvili.budko.07}.

The existence of the topological phase is indicated by interesting transport and magnetic properties of LaAgSb$_{2}$. 
For example, a large linear magnetoresistance and a positive Hall resistivity were observed~\cite{,myers.budko.99,myers.budko.99b,wang.petrovic.12}.
Similar behavior was also found in the compound with the same structure, namely the topological magnetic system SrMnBi$_{2}$~\mbox{\cite{park.lee.11,wang.graf.11,wang.zhao.11}}.
In that case, the transport behavior is related to the occurrence of the anisotropic Dirac states, where linear energy dispersion originates from the crossing of the $p$-orbital bands in the double-size Bi square net~\cite{park.lee.11}.
The first principles study indicated similar properties in the case of LaAgSb$_{2}$, where the $p$ orbitals of Sb atoms create the bands with a nearly-linear dispersion~\cite{wang.petrovic.12}.
More recently, the angle-resolved photoemission spectroscopy (ARPES) of LaAgSb$_{2}$ was performed~\cite{shi.richard.16}.

\paragraph*{Motivation.---}

The aforementioned ARPES~\cite{shi.richard.16} study indicates the existence of several anomalies in the spectral function. In particular, the Dirac-cone-like structure, which originates from the band crossing, was found along the $\Gamma$--M direction.
In this paper, we performed systematic and complex study of the electronic band structure of LaAgSb$_{2}$, using both high quality ARPES measurements and theoretical analyzes.
The theoretical study of the electronic band structure was based on the density functional theory (DFT) calculations, as well as on the tight binding model. 
Possible surface states were predicted within the direct DFT calculation for slab geometry and the surface Green function technique for the (001) surface.
We show that the realized surface states strongly depend on the termination of the system.
Theoretically predicted surface states at $\Gamma$ and X were confirmed by the ARPES measurements. 
A presence of nodal lines, which originate either from lattice symmetry or vanishing of spin-orbit splitting, is discussed. 
Additionally, we analyze the possible Dirac surface states along the $\Gamma$--M direction~\cite{shi.richard.16}. 
Our theoretical and experimental study does not confirm realization of such states along this direction.

This paper is organized as follows. 
Experimental and theoretical methods are briefly described in Sec.~\ref{sec.tech}.
Next, we presented and discuss our results in Sec.~\ref{sec.res}.
Finally, Section~\ref{sec.summary} contains conclusions and a summary.


\section{Techniques}
\label{sec.tech}

\subsection{Sample preparation}

Single crystals of LaAgSb$_{2}$ were grown by the self-flux technique similar to the method reported in Ref.~\cite{myers.budko.99b} and also applied for the growth of UCoSb$_{2}$ crystals~\cite{bukowski.tran.04}.
Lanthanum (purity $99.9$\%), silver (purity $99.99$\%), and antimony (purity $99.999$\%) were used as starting materials.
The components were weighed in the atomic ratio La:Ag:Sb = 1:1.1:22 and placed in an alumina crucible, which was then sealed in an evacuated silica tube.
The ampoule was heated at $1100$~$^{\circ}$C for $5$~h followed by slow cooling ($2$~$^{\circ}$C/h) down to $680$~$^{\circ}$C.
At this temperature, the ampoule was flipped upside down in order to decant still liquid Sb-Ag flux.
Next, the ampule was fast cooled to room temperature and the crucible was transferred to another silica tube, where the rest of Sb was removed from crystals by means of sublimation in high vacuum at $600$~$^{\circ}$C.
Finally, single crystals were mechanically isolated.

The obtained crystals were examined using a scanning electron microscope (SEM) and their chemical composition was determined with energy dispersive X-ray spectroscopy (EDX)  using a standardless procedure.
The crystal structure and lattice parameters were determined using powder X-ray diffraction on a sample of the crushed single crystal.

\subsection{Experimental details}

High resolution angle resolved photoemission studies were performed using two ARPES systems, one at Solaris Synchrotron, Krak\'{o}w, Poland, at UARPES beamline,  equipped with Scienta-Omicron DA30-L electron analyzer, and the other at our in-house laboratory with Scienta R4000 analyzer using He-I radiation ($h\nu = 21.218$~eV).
Samples measured with use of the synchrotron radiation were cleaved in situ in ultrahigh vacuum at room temperature and the measurements were performed at the temperature of $12$~K, while in the in-house laboratory the samples were cleaved and measured at a temperature of $20$~K. 
Base pressure was below $5 \times 10^{-11}$ mbar in both systems.
In the case of measurements performed at the synchrotron, the spectra were collected in the vertical and horizontal polarization of the incident light. 
The ARPES results are presented as the sum of these two scans in order to reduce matrix element contribution. 
In order to comprehensively analyze the obtained data, the 2D curvature method~\cite{zhang.richard.11}  was used.

\begin{figure*}
\includegraphics[width=\linewidth]{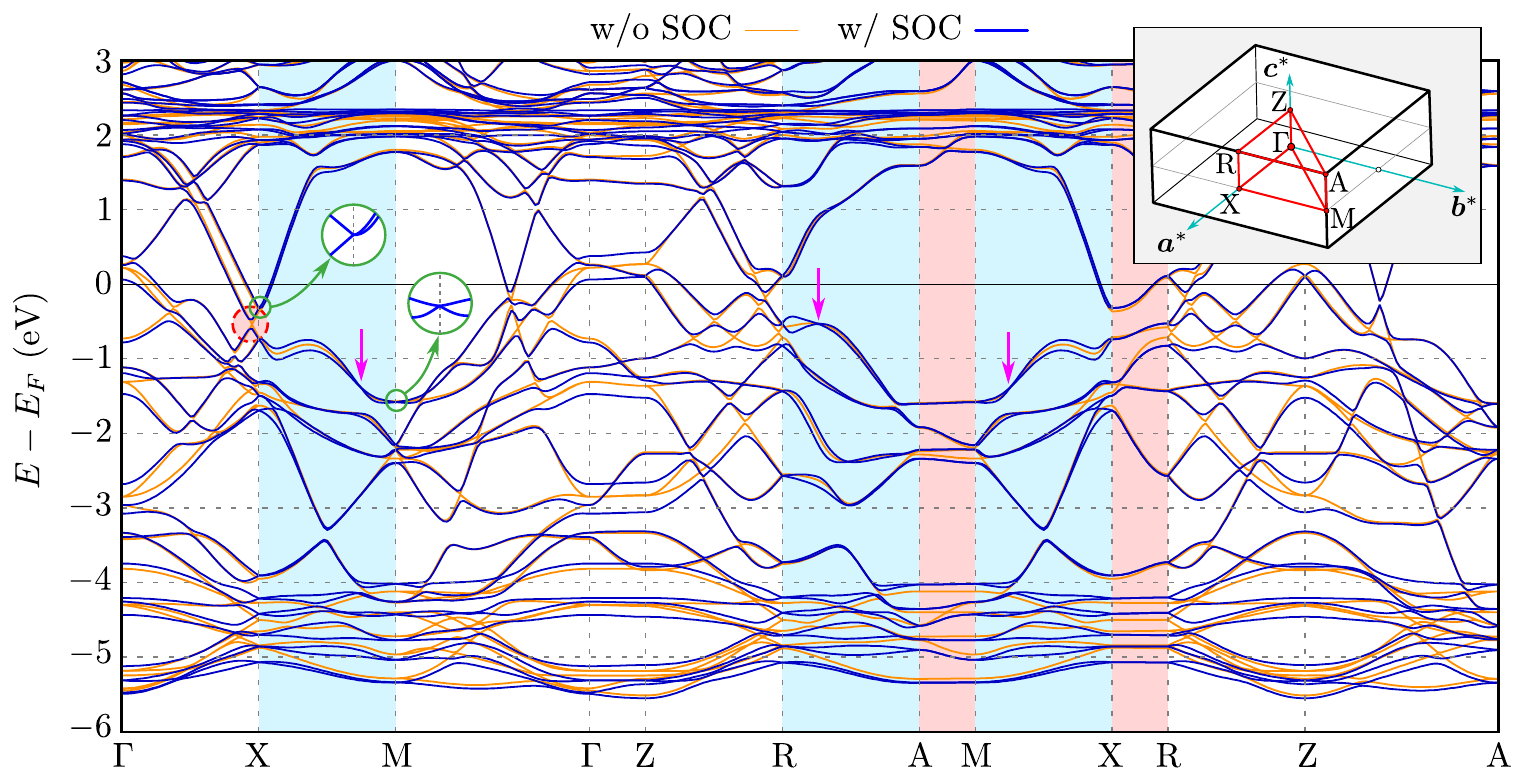}
\caption{
The bulk band structure in the absence and presence of the spin--orbit coupling (orange and blue lines, respectively). 
X--M and R--A paths, where nodal lines can be realized due to vanishing spin--orbit coupling, are marked with blue background.
Similarly, X--R and A--M paths are highlighted with red background, where nodal lines with fourfold degeneration exist due to the {\it P4/nmm} symmetry.
Two circular insets highlight the bulk Dirac cones at the X and M points.
Top right inset presents the Brillouin zone of the tetragonal structure {\it P4/nmm} and its high-symmetry points~\cite{setyawan.curtarolo.10}. }
\label{fig.bulk_bands}
\end{figure*}

\subsection{Calculation details}

The DFT calculations were performed within the projector augmented-wave (PAW) method~\cite{blochl.94} using the Vienna 
\textit{Ab initio} Simulation Package ({\sc vasp})~\cite{kresse.hafner.94,kresse.furthmuller.96,kresse.joubert.99}.
The exchange-correlation potential was obtained by the generalized 
gradient approximation (GGA) in the form proposed by Perdew, Burke, and Enzerhof (PBE)~\cite{pardew.burke.96}. 
We also investigated the impact of the spin--orbit coupling (SOC)~\cite{steiner.khmelevskyi.16} on the electronic structure.

The optimization of both the structural parameters and the electronic structure was performed using a $20\times 20\times 10$ Monkhorst-Pack {\bf k}-grid~\cite{monkhorst.pack.76}. 
The energy cut-off for the plane-wave expansion was equal to $520$~eV.
The structures were relaxed using the conjugate gradient technique with the energy convergence criteria set at $10^{-8}$~eV and $10^{-6}$~eV for the electronic and ionic iterations, respectively.
Next, the optimized structure was used to construct of the tight binding model in the maximally localized Wannier orbitals (cf.~Sec.~\ref{sec.model_tbm}).
The surface Green function for semi-infinite system~\cite{sancho.sancho.85} for study the surface states was calculated using {\sc WannierTools}~\cite{wu.zhang.18}.


\section{Results and discussion}
\label{sec.res}

\subsection{Crystal structure}
\label{sec.crys_stru}

LaAgSb$_{2}$ crystallizes in the tetragonal ZrCuSi$_{2}$-type lattice~\cite{thirion.venturini.83}, with the layered structure {\it P4/nmm} (space group 129) presented in Fig.~\ref{fig.crys}.
The obtained experimental crystals are rectangular and platelike with a crystallographic $c$ axis perpendicular to the plane.
The chemical composition of the crystals, determined from the EDX data, corresponded well to the ideal formula LaAgSb$_{2}$.
The lattice parameters determined from XRD ($a = b = 4.390$~\AA\ and $c = 10.84$~\AA) are in good agreement with those reported in the literature ($a = b = 4.359$~\AA, and $c = 10.787$~\AA~\cite{brylak.moller.95,gondek.penc.02}).
In our calculations, the lattice constants were found as $a = b = 4.43974$~\AA, and $c = 10.89658$~\AA, which are in good agreement with the experimental results.
The Wyckoff positions of atoms are La: $2c$ ($\sfrac{1}{4}$,$\sfrac{1}{4}$,$0.73994$), Ag: $2a$ ($\sfrac{3}{4}$,$\sfrac{1}{4}$,$0$), Sb(1): $2b$ ($\sfrac{3}{4}$,$\sfrac{1}{4}$,$\sfrac{1}{2}$), and Sb(2): $2c$ ($\sfrac{1}{4}$,$\sfrac{1}{4}$,$0.17137$).
The Sb(1) atoms form a two-dimensional square lattice, while the Sb(2) atoms coordinated by La atoms form a LaSb(2) double layer characterized by the glide symmetry. The Ag atoms reside between two LaSb(2) layers.
The layered structure can be viewed as a sequential stacking of Sb(1)-LaSb(2)-Ag-LaSb(2)-Sb(1) layers along the $c$ axis (cf.~Fig.~\ref{fig.crys}).

\subsection{Band structure and Fermi surface}
\label{sec.bulkband}

The {\it ab inito} (DFT) calculated bulk band structure is presented in Fig.~\ref{fig.bulk_bands}, where we show 
the comparison of the results in the absence and presence of the spin--orbit coupling (gray and blue lines respectively). 
They are in agreement with the previous study~\cite{wang.petrovic.12,hase.yanagisawa.14,ruszala.winiarski.20}.
The unoccupied La 5$f$ orbitals are located above the Fermi level (around $2$~eV).
The 5$p$ orbitals are responsible for the emergence of the nearly-linear band crossing the Fermi level~\cite{wang.petrovic.12,ruszala.winiarski.20}.
These bands are associated with the several places where band crossing exists.
These points arise due to the folding of the dispersion relation of the $p$-orbitals in Ag and Sb nets.
Including the spin--orbit coupling removes the degeneracy of these points.

At the X point, we can also observe the reduction of the irreducible representations of bands due to including the SOC.
Without the SOC, there are four different irreducible representations of bands, while the presence of the SOC allows for only one irreducible representation and causes band splitting and gap opening close to the X point (marked by the dashed red circle in Fig.~\ref{fig.bulk_bands}).
Additionally, the energy difference between two Dirac-like cones at the X point (first zoom inside left inset in Fig.~\ref{fig.bulk_bands} and second non-shown below them) typically strongly depends on $c/a$~\cite{topp.lippmann.16}.
The described modification originates mostly from folding of the band structures of Ag and Sb square nets~\cite{klemenz.schoop.20}, due to realization of their double unit cell, i.e. $\sqrt{2} \times \sqrt{2} \times 1$ cell, within the LaAgSb$_{2}$ cell 
[cf.~Fig.~\ref{fig.bands_layer} in Supplemental Material (SM)~\footnote{See Supplemental Material at [URL will be inserted by publisher] for additional theoretical results.}].


\paragraph*{Highly degenerated nodal lines.---}
Including SOC leads to lifting of the band degeneracy, while the time reversal symmetry is still preserved due to the absence of the magnetic order.
Additionally, the nonsymmorphic glide symmetry $\{ 2_{110} | \sfrac{1}{2},\sfrac{1}{2},0\}$, mapping $( k_{x} , k_{y} , k_{z} ) \rightarrow ( -k_{y} , -k_{x} , -k_{z} )$, is guaranteed by the space group {\it P4/nmm}.
Assembling these symmetries with the three-dimensional (3D) inversion symmetry has a consequence in the form of the bulk band structure~\cite{young.kane.15}.
Along the high symmetry directions, all the bands at the X and M points are doubly degenerated by the glide mirror symmetry, forming Dirac cones (two left insets in Fig.~\ref{fig.bulk_bands} and Fig.~\ref{fig.bulk_dirac} in SM~\cite{Note1}).
Extending to 3D~\cite{young.kane.15}, all lines along $k_{z}$ containing X and M points (i.e. X--R and M--A) form degenerated lines called nodal lines (NLs).
NLs are formed from the crossing of two Dirac-like bands. 
Similar situation is observed in the other Dirac semimetals~\cite{fu.yi.19,chen.xu.17,takane.wang.16,hosen.dimitri.17,schoop.ali.16,neupane.belopolski.16,topp.queiroz.17,schoop.topp.18,regmi.dhakal.21,topp.lippmann.16,schoop.ali.16,neupane.belopolski.16,lou.ma.16,hosen.dimitri.18,wang.qian.21}.

\begin{figure}[!b]
\includegraphics[width=\linewidth]{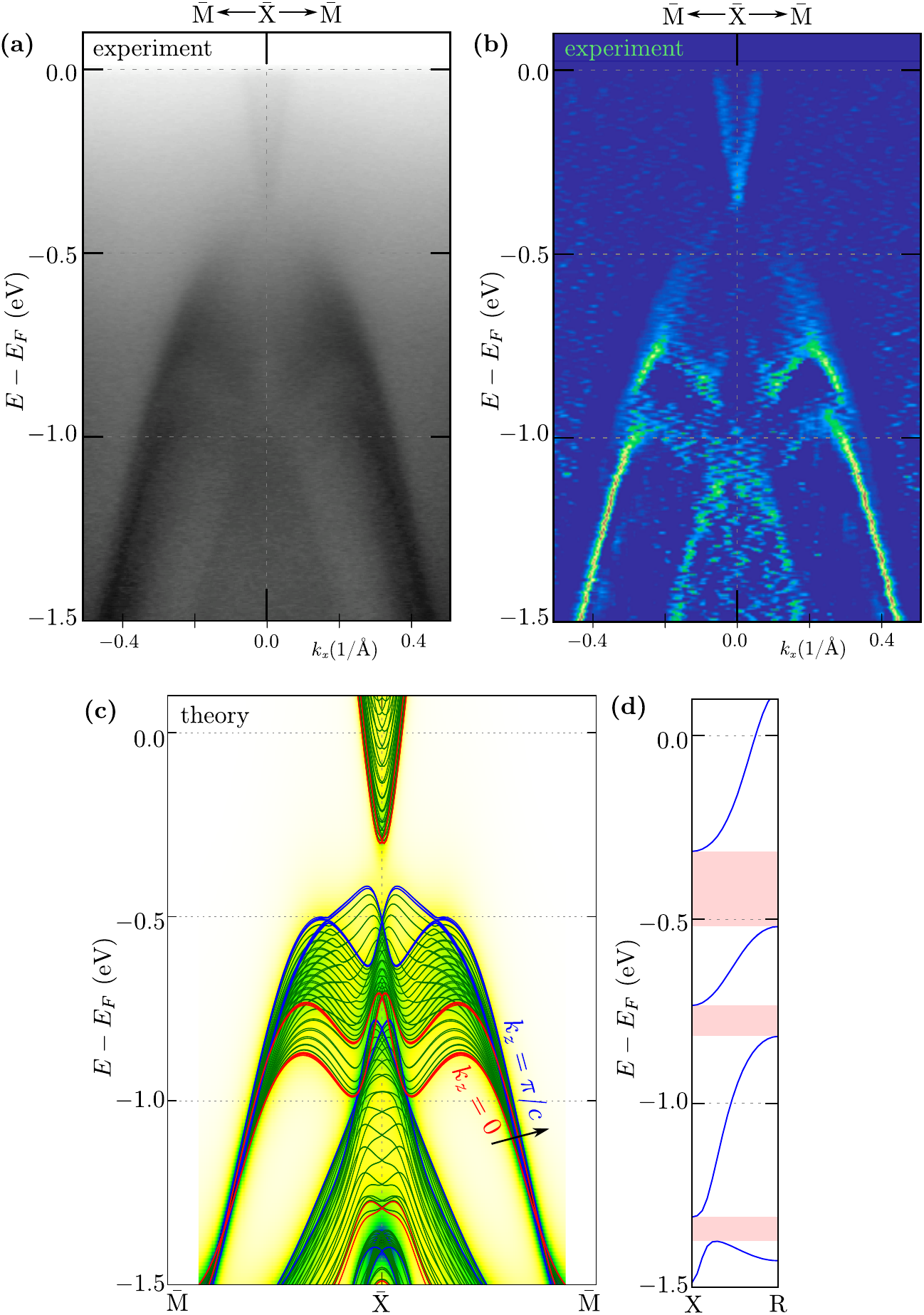}
\caption{
Comparison of the ARPES spectra (a) and corresponding 2D curvature (b) with the theoretical calculation: spectral function (background) and projected bulk bands (lines) (c).
The results are presented along the $\bar{\text{X}}$--$\bar{\text{M}}$ direction, around the $\bar{\text{X}}$ point. 
The experimental data have been collected with photon energy $h\nu = 66$~eV at a temperature of $12$~K. 
Panel (d) presents the energy range of the nodal lines in the bulk band structure along the X-R path (red areas mark band gap between the nodal lines). 
}
\label{fig.arpes_cut_x}
\end{figure}

In a general case, NLs can be observed experimentally, in a form of the band crossing~\cite{song.wang.20}. 
Its binding energy varies with $k_{z}$, what can be visualized by changing photon energy.
In our case, NLs should be visible in the form of the band crossing at the $\bar{\text{X}}$ points, where the bulk NLs from X--R path are projected (Fig.~\ref{fig.arpes_cut_x}).
Theoretical investigation of the bulk NLs recovers existence of several gaps between them [marked by red areas in Fig.~\ref{fig.arpes_cut_x}(d)].
For instance, NLs inside the upper band fill approximately the energy range from $-0.7$~eV to $-0.5$~eV below the Fermi level.
In this range of energies, several $k_{z}$-depending band crossings are visible [see Fig.~\ref{fig.arpes_cut_x}(c)]
Unfortunately, the exact band crossing is not well resolved in the experiment (top panels on Fig.~\ref{fig.arpes_cut_x}).
However, depletion of spectral intensity in the experiment is observed for binding energies around $\sim-0.4$~eV corresponding to the gaps at $\bar{\text{X}}$. 
Additionally, we observed excellent agreement of the experimental results with the theoretical ones (cf.~top and bottom bands in Fig.~\ref{fig.arpes_cut_x}).

\begin{figure}[!t]
\includegraphics[width=\linewidth]{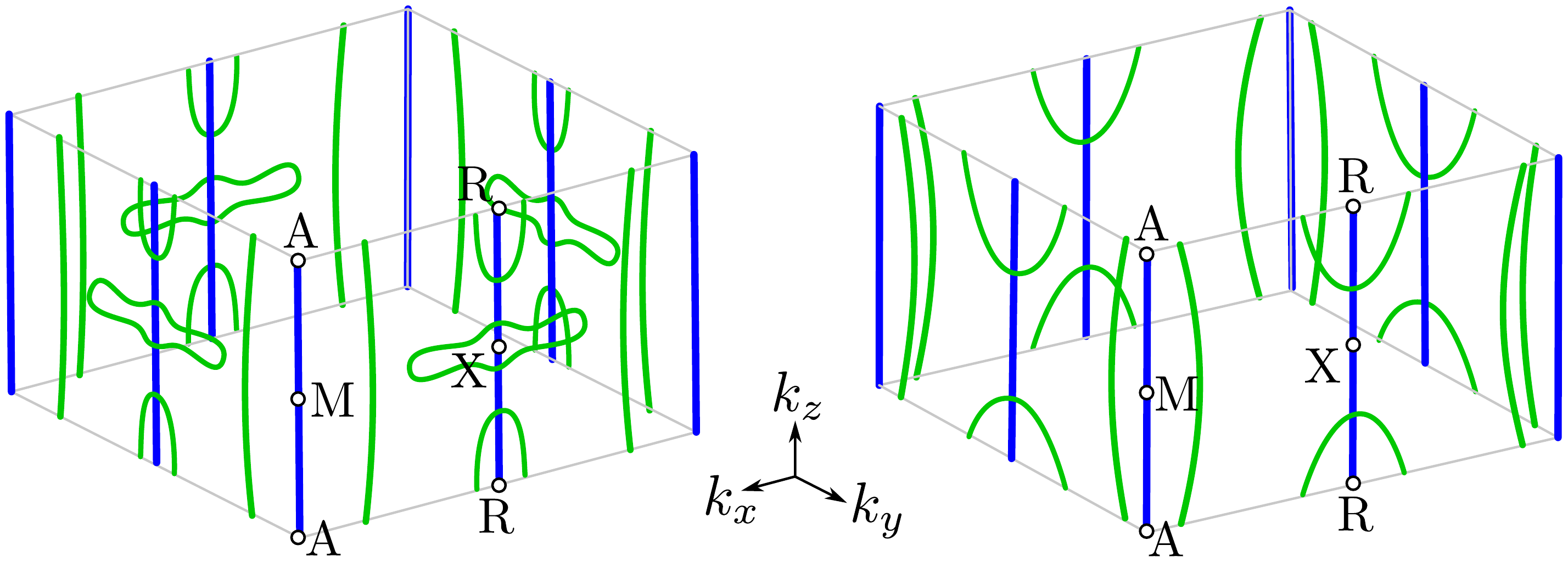}
\caption{
Brillouin zone with highly degenerated nodal lines in the electronic band structure realized due to the symmetry of the system (blue lines) or as a consequence of vanishing spin--orbit coupling on the Brillouin zone boundary (green lines). 
Presented green curves are realized by the nodal lines with energies from ranges $-1$~eV to $-2$~eV, and from $-0.5$ to $-1.0$~eV, for left and right panel, respectively.
}
\label{fig.dnl_schem}
\end{figure}

The SOC slightly lifts the degeneracy along the X--M and R--A lines.
However, the highly degenerated nodal lines (e.g., places marked by pink arrows in Fig.~\ref{fig.bulk_bands}) also can be found in this direction, which is due to vanishing of the SOC for some ${\bm k}$.
These fourfold degenerated Dirac points lead to the emergence of an additional NL at the boundary of the Brillouin zone (in the X--M--A--R planes).
The NLs are presented schematically in Fig.~\ref{fig.dnl_schem}, by green lines (we present NLs only for eight bands near the Fermi level). 
As we can see, some of the NLs create closed contours along $k_{z}$ located near the A--M edge of Brillouin zone.
Contrary to this, other NLs form closed ellipsoidal contours around R high symmetry point.

\begin{figure}[!b]
\includegraphics[width=0.6\linewidth]{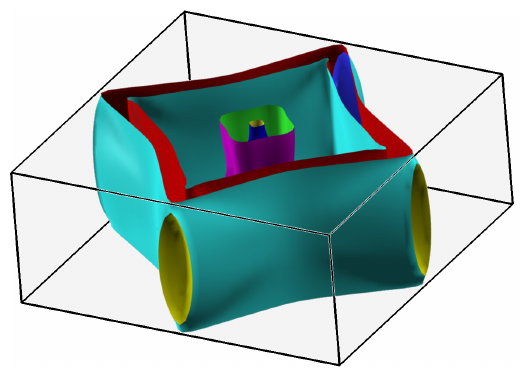}
\caption{
The Fermi surface of LaAgSb$_{2}$ in the absence of the spin--orbit coupling.
}
\label{fig.fbz}
\end{figure}

\begin{figure*}[!t]
\includegraphics[width=\linewidth]{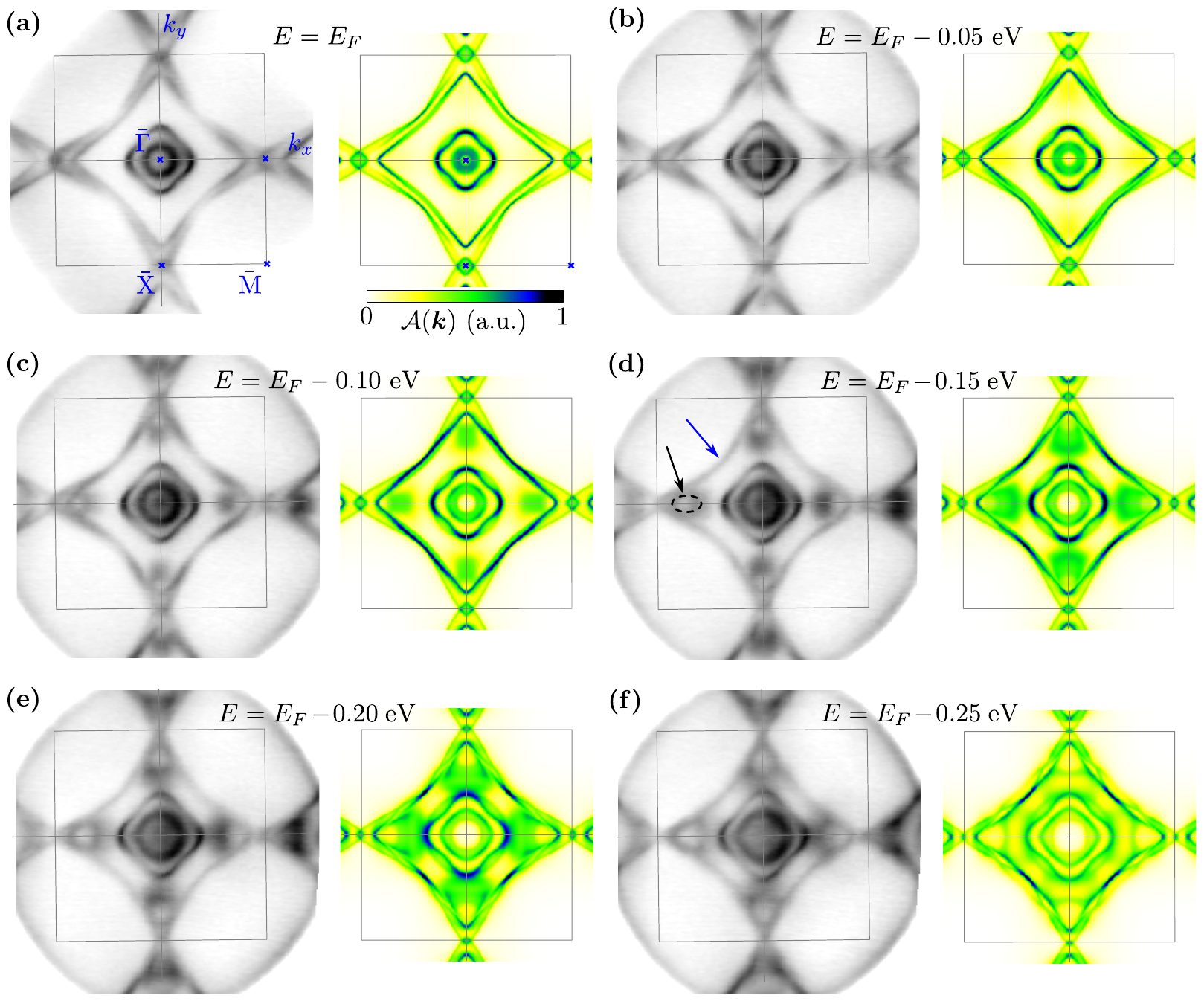}
\caption{
Comparison of the ARPES spectra obtained with a photon energy of $66$~eV at a temperature of $12$~K (left panels) with the theoretical surface spectral function (right panels) for selected binding energies $E$ below the Fermi level (as labeled).
The gray square denotes the Brillouin zone. 
Black arrow and the dashed outline indicate the additional branch explained by the spectral function.
\label{fig.arpes_cute}
}
\end{figure*}

\begin{figure*}
\includegraphics[width=\linewidth]{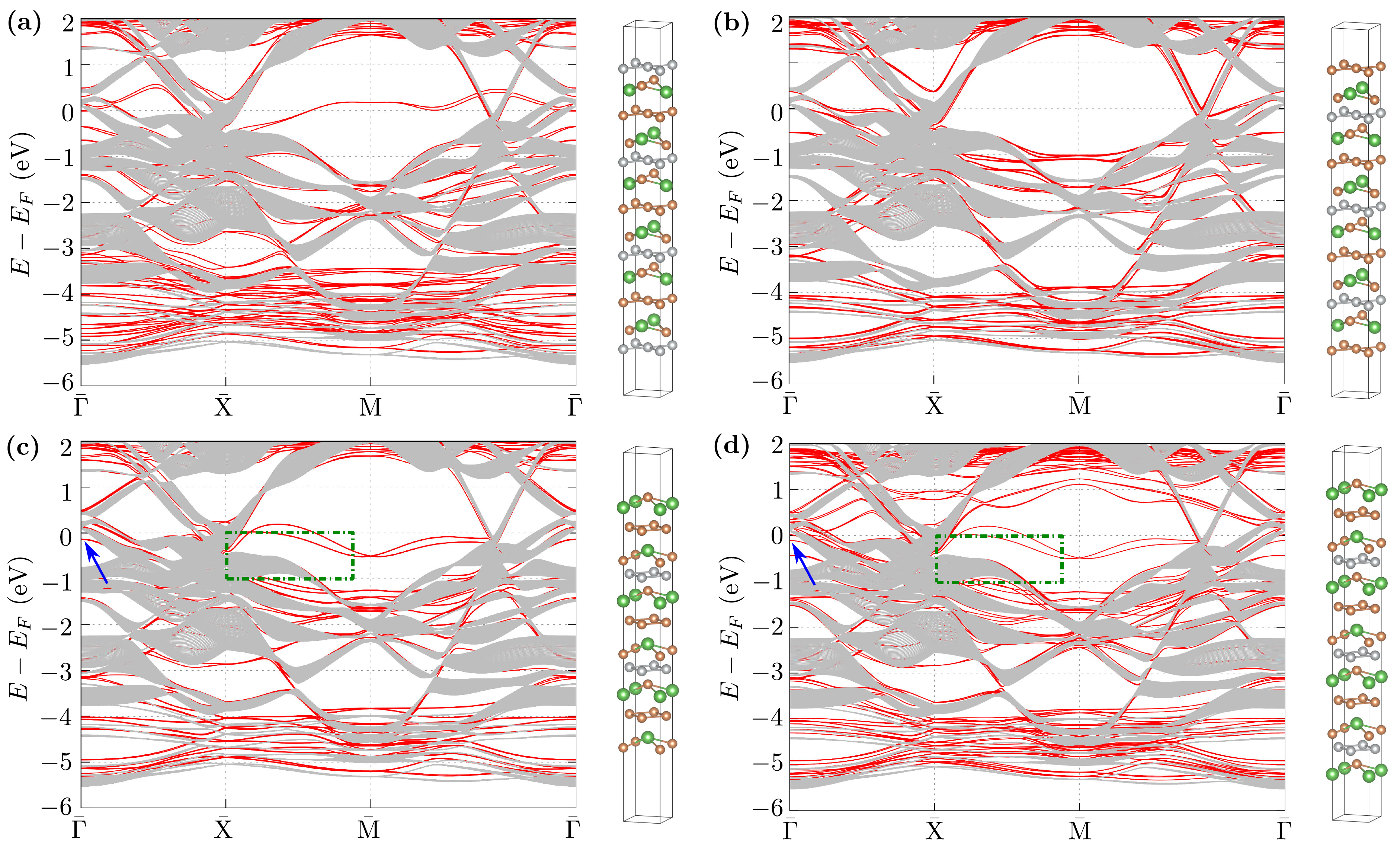}
\caption{Band structures in slab geometry (with corresponding structures presented on right hand side) dependent on the surface termination of Ag (a) and Sb (b) square nets, and two possible terminations of LaSb layer -- conserving (c) and breaking (d) glide symmetry.
Red lines denote slab band structure from the DFT calculations, while gray lines denote projected bulk band structure from tight binding models calculations (for eighty $k_{z}$ different momenta).
Blue arrows indicate the surface states below the Fermi level at $\Gamma$ point.
Green dotted-dashed line in (c) and (d) shows the area presented in Fig.~\ref{fig.arpes_ss}.
}
\label{fig.bands_slab}
\end{figure*}

\paragraph*{Fermi surface.---}
The previous studies showed that the FS in the absence of the SOC (Fig.~\ref{fig.fbz}) is composed of four bands~\cite{myers.budko.99,hase.yanagisawa.14,akiba.umeshita.22,bosak.souliou.21}: two cylindrical-like pockets along the $\Gamma$--Z direction and two pockets with diamond-like shape connecting the X points of the first Brillouin zone~(separate pockets are shown in Fig.~\ref{fig.fs_com} in SM~\cite{Note1}).
Analyses of the band structure of the separate layers of LaAgSb$_{2}$, i.e. Sb and Ag square nets, and LaSb double-layer (cf. Fig.~\ref{fig.bands_layer} in SM~\cite{Note1}), clearly show that the pockets around the $\Gamma$--Z direction are associated with the LaSb double layer (mostly 5$d$ orbitals of La and 5$p$ orbitals of Sb).
Similarly, the quasi-two dimensional Sb and Ag square nets create diamond-like pockets (5$s$ and 4$d$ orbitals of Ag and 5$p$ orbitals of Sb).

The isoenergetic study of the spectral function clearly shows that the characteristic shape of the FS is observed experimentally. 
A comparison of the experimental and theoretical results is presented in Fig.~\ref{fig.arpes_cute}, left and right panels, respectively.
First, the ARPES result for the energy corresponding to the Fermi level $E_{F}$ [Fig.~\ref{fig.arpes_cute}(a)] is clearly reproduced by the theoretical projected Fermi surface. 
Similarly as in the bulk system [Fig.~\ref{fig.fbz}(b)], we can find two pockets centered at the $\bar{\Gamma}$ point and two diamond-like pockets connecting the $\bar{\text{X}}$ points.
Our detection of the pocket around the $\bar{\Gamma}$ point is in agreement with the previous measurements for this material~\cite{arakane.sato.07}.
In both the experimental and theoretical results, we can find a few characteristic features.
The internal $\bar{\Gamma}$-centered pocket shows weak $k_{z}$-dependence (theoretical spectral weight of this pocket has an approximately constant intensity).
Contrary to this, the external $\bar{\Gamma}$-centered pocket shows stronger $k_{z}$-dependence in the $\bar{\Gamma}$--$\bar{\text{M}}$ direction than in $\bar{\Gamma}$--$\bar{\text{X}}$ -- this is well visible in a form of the increasing theoretical spectral weight along the second direction. 
The situation looks similar in the case of the diamond-like pockets. 
Here, we must mention that the similar diamond-like FS projection is observed also in the other Dirac semimetals within this family, e.g. LaCuSb$_{2}$~\cite{chamorro.topp.19}, YbMnBi$_{2}$~\cite{borisenko.evtushinsky.19} and YbMnSb$_{2}$~\cite{kealhofer.jang.18}, or SrMnBi$_{2}$ and CaMnBi$_{2}$~\cite{feng.wang.14}, as well as a square-net materials~\cite{hosen.dimitri.17}, e.g. ZrSiTe~\cite{topp.lippmann.16}, \mbox{ZeSiS}~\cite{schoop.ali.16,neupane.belopolski.16,topp.queiroz.17}, ZrSnTe~\cite{lou.ma.16}, ZrGeTe~\cite{hosen.dimitri.18}, HfSiS~\cite{takane.wang.16,chen.xu.17}, ZrSiS~\cite{chen.xu.17,fu.yi.19}, SmSbTe~\cite{regmi.dhakal.21}, LaSbTe~\cite{wang.qian.21}, or NbGeSb~\cite{markovic.hooley.19}.

The characteristic features for $E = E_{F}$ are conserved also below $E_{F}$.
However, with going below $E = E_{F}$ we can observe a few important behaviors.
First, we notice $E$ dependence of two branches forming two diamond-like pockets of the FS [marked with blue arrow at Fig.~\ref{fig.arpes_cute}(d)]. 
Under decreasing $E$ we find the hourglass-shape by the band branches (what will be discussed more precisely in the next paragraphs).
Secondly, around $E = E_{F} - 0.1$~eV we can observe the appearance of the parabolic-like branch in the band structure near the $\bar{\text{X}}$ point.
For lower $E$, this branch is well visible in the form of a circular shape in the spectral function [marked by black dashed lines and black arrow in Fig.~\ref{fig.arpes_cute}(d)]. 
For $E = E_{F} - 0.25$~eV this branch is linked to the diamond-like shape.
It is worth mentioning that in the whole range of $E$, the theoretical calculations are comparable with the experimental ARPES results [cf. left and right panels in Fig.~\ref{fig.arpes_cute}(a)--(f)], which indicate a weak role of the electron correlations in this system in the considered energy scale.

With decreasing $E$, modification of the FS pocket around the $\bar{\text{X}}$ point can be observed (Fig.~\ref{fig.arpes_cute}).
This pocket is formed by a common part of two diamond-like pockets from the two nearest Brillouin zones.
The change in the size of these pockets observed for the decreasing binding energy leads to the disappearance of the pocket crossing the border of the Brillouin zone [cf. panels from Fig.~\ref{fig.arpes_cute}(a) to Fig.~\ref{fig.arpes_cute}(f)].
Indeed, we can reveal this behavior carefully analyzing both experimental data and the theoretical spectral function along the $\bar{\text{X}}$--$\bar{\text{M}}$ direction, i.e. at the Brillouin zone boundary [Fig.~\ref{fig.arpes_cut_x}(c)]. 
The electron pocket crossing the Brillouin zone boundary vanishes around the energy $\sim 0.35$~eV below the Fermi level. Theoretical analyses show, that the upper (electron) (above $\sim -0.35$~eV) and lower (hole) bands (below $\sim -0.5$~eV) are separated at the $\bar{\text{X}}$ point by the $\sim 0.15$~eV band gap.
The lower bands create a characteristic double parabolic structure corresponding to experimental results [Fig.~\ref{fig.arpes_cut_x}(a)].

\subsection{Surface states within direct DFT slab calculations}
\label{sec.slabdft}

To evaluate the data described in the previous section, we performed the DFT calculation for slab geometry with different terminations (Fig.~\ref{fig.bands_slab}), namely square net of Ag atoms [Fig.~\ref{fig.bands_slab}(a)], square net of Sb atoms [Fig.~\ref{fig.bands_slab}(b)], and two nonequivalent LaSb layers shown in [Fig.~\ref{fig.bands_slab}(c) and Fig.~\ref{fig.bands_slab}(d)]. 
The structural models used in the DFT calculations are presented at the right side of the obtained band structures [Fig.~\ref{fig.bands_slab})]. 
Each time $\sim 10$~\AA\ of vacuum is included.

To extract the surface states, we compare calculated band structures of slabs (solid red lines), with the bulk band structures (solid grey lines) projected onto the 2D reduced Brillouin zone.
In the case of the bulk bands, around the $\bar{\text{M}}$ point there exists a relatively large ``gap'' between continuum states which allows in a relatively simple way to extract surface states. Indeed, in this region the surface states are well visible. 
For three cases of chosen terminations, the surface states crossing the Fermi level are well visible around the M point, where they form Dirac cones located either above [Fig.~\ref{fig.bands_slab}(a)] or below [Fig.~\ref{fig.bands_slab}(c), and Fig.~\ref{fig.bands_slab}(d)] $E_F$. 
In the case of Sb termination [Fig.~\ref{fig.bands_slab}(b)], the surface states are deep below the Fermi level (around $-1$~eV).
Similar states can be observed experimentally in some materials, like NbGeSb~\cite{markovic.hooley.19}. 
Absence of such Dirac surface states at M in the photoemission data would suggest the realization of the Sb or Ag square net termination. Similarly, only LaSb termination allows the realization of the surface states near $E_F$ in the region of $\Gamma$ point, shown by blue arrows in Fig.~\ref{fig.bands_slab}(c) and Fig.~\ref{fig.bands_slab}(d).
Nevertheless, the most interesting range of energies (around the Fermi level) corresponding to the paths $\bar{\Gamma}$--$\bar{\text{X}}$ and $\bar{\text{M}}$--$\bar{\Gamma}$ is filled by the bulk states.
Contrary, e.g., to the topological insulators~\cite{hsieh.xia.09,zhang.liu.09,xia.qian.09,alpichshev.analytis.10,kuroda.arita.10}, the extraction of separate surface states can be hard or even impossible within the existing ARPES data due to the presence of bulk bands.

\paragraph*{Role of surface termination.---}
As we can see from the comparison of the results for different surface terminations, there exists a strong dependence of the realized surface states on a type of termination (cf. panels on Fig.~\ref{fig.bands_slab}).
Such behavior was reported before, e.g., in the case of Weyl semimetals~\cite{su.wu.15}.
At the same time, in each case, nearly linear bulk bands crossing the Fermi level are insensitive to modification of termination.
The same bulk bands play the most important role in the described earlier ARPES data, due to the weak $k_z$ dependence of the corresponding dispersion relations.
To conclude this part, we can propose that the sample studied experimentally was terminated by the LaSb layer.
However, the discussed surface states should be absent, if the surface termination resulting from cleaving is different.

\subsection{Surface states around $\bar{\Gamma}$ and $\bar{\text{X}}$}
\label{sec.surf_state_gx}

\begin{figure}[!b]
\includegraphics[width=\linewidth]{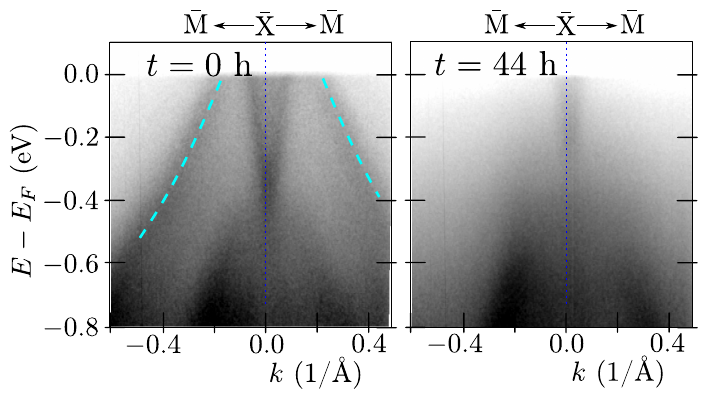}
\caption{
The ARPES results obtained around the $\bar{\text{X}}$ point presenting the realized surface states. The panels present spectra collected at different time following the sample cleaving. 
Cyan dashed lines highlight the surface state.
The measurements have been realized with He-I ($h\nu = 21.218$~eV) radiation for the sample cleaved and measured at a temperature of~$20$~K 
}
\label{fig.arpes_ss}
\end{figure}

ARPES measurements, which visualize surface states better, were performed directly following the cleaving at $20$~K.
For this case, the ARPES spectra obtained along $\bar{\text{X}}$-$\bar{\text{M}}$ direction, around $\bar{\text{X}}$ point are presented in Fig.~\ref{fig.arpes_ss}.
Now, we observe an additional band in the range of energies, where bulk states are not realized.
This additional band disappeared much quicker compared to the rest of the structure (cf. left and right panels, presenting results at a different time following the cleaving of the crystal).
What is important, the decay time was different in different vacuum systems, namely from several minutes to tens of hours.
This is a consequence of surface contamination by impurities from the environment.
All this makes a strong suggestion that the described band is of surface origin.
By comparing these results with the theoretical prediction, we can conclude that the LaSb termination is realized for this sample [cf.~Fig.~\ref{fig.bands_slab}(c) and Fig.~\ref{fig.bands_slab}(d)].

\begin{figure*}[!pt]
\includegraphics[width=0.85\linewidth]{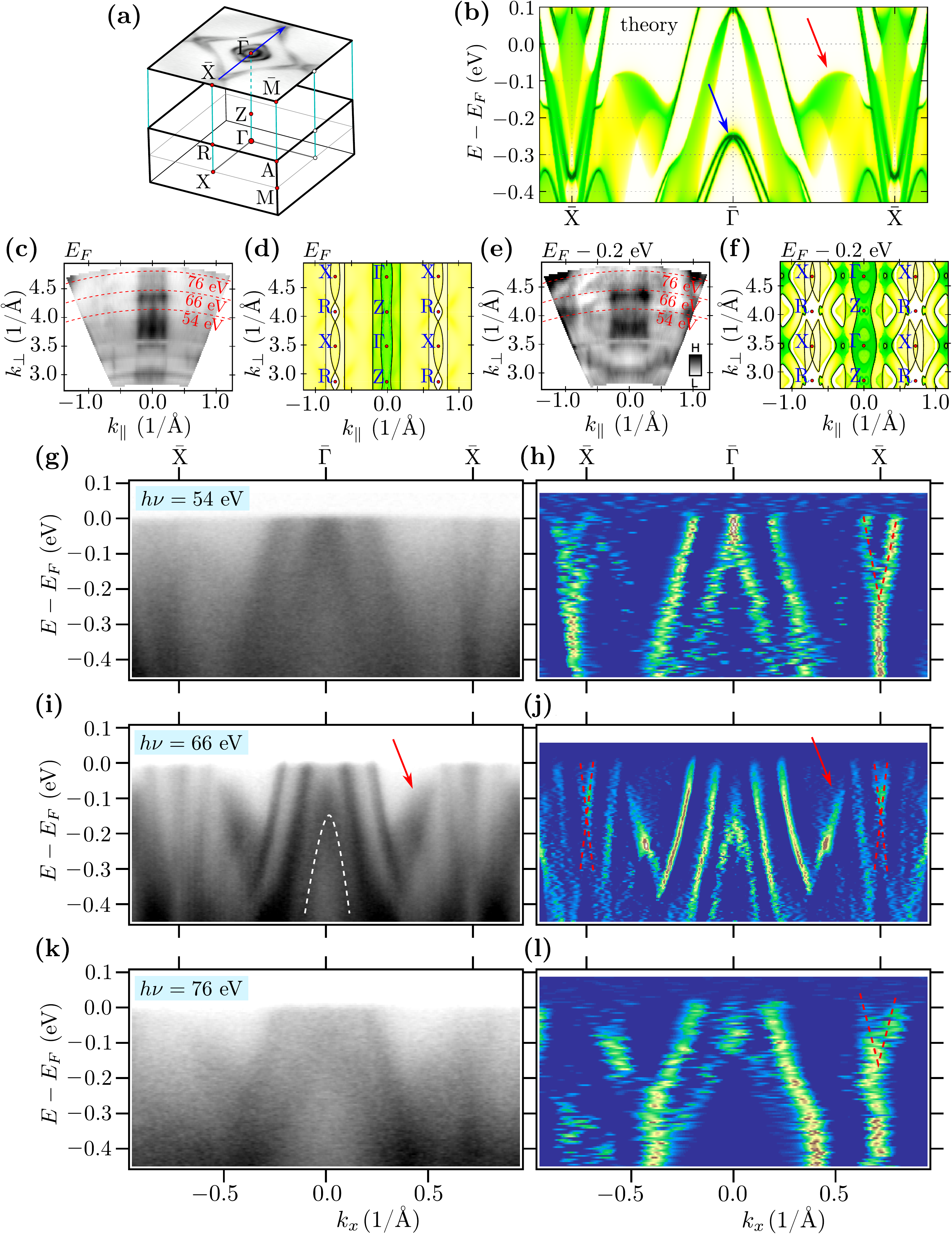}
\caption{
Cartoon representation of the bulk first Brillouin zone projection into the surface Brillouin zone (a).
Theoretical surface spectral function along the $\bar{\text{X}}$--$\bar{\Gamma}$--$\bar{\text{X}}$ direction (b).
Panels from (c) to (f) present intensity plots for constant energy planes at the Fermi level $E_{F}$ and energy $E_{F}-0.2$~eV (as labeled).
The ARPES spectra (c) and (e) were collected in the photon energy range from $20$~eV to $80$~eV. 
The calculated band structure (solid black lines) along with the spectral function (background color scale) is shown in (d) and (f).
Here the intensity maps is a function of $k_{\perp} = k_{z}$ and $k_{\parallel} = k_{x}$ (perpendicular and parallel to the surface component of the wave vector).
Panels from (g) to (l) present experimentally obtained spectra and their 2D curvature (left and right panels, respectively), for different photon energies (as labeled). 
Red arrows in (b), (i) and (j) indicate the position of additional structure in the spectra that can be explained by the projection of the bulk  band on the surface Brillouin zone. 
Blue arrow in (b) and white dashed parabola in (i) indicate the surface state. Red dashed lines in (h), (j) and (l) refer to the position of the band crossing of the band forming the nodal line.
The measurements were made at the temperature of $12$~K.
}
\label{fig.dnl1}
\end{figure*}

\begin{figure*}[!pt]
\includegraphics[width=0.95\linewidth]{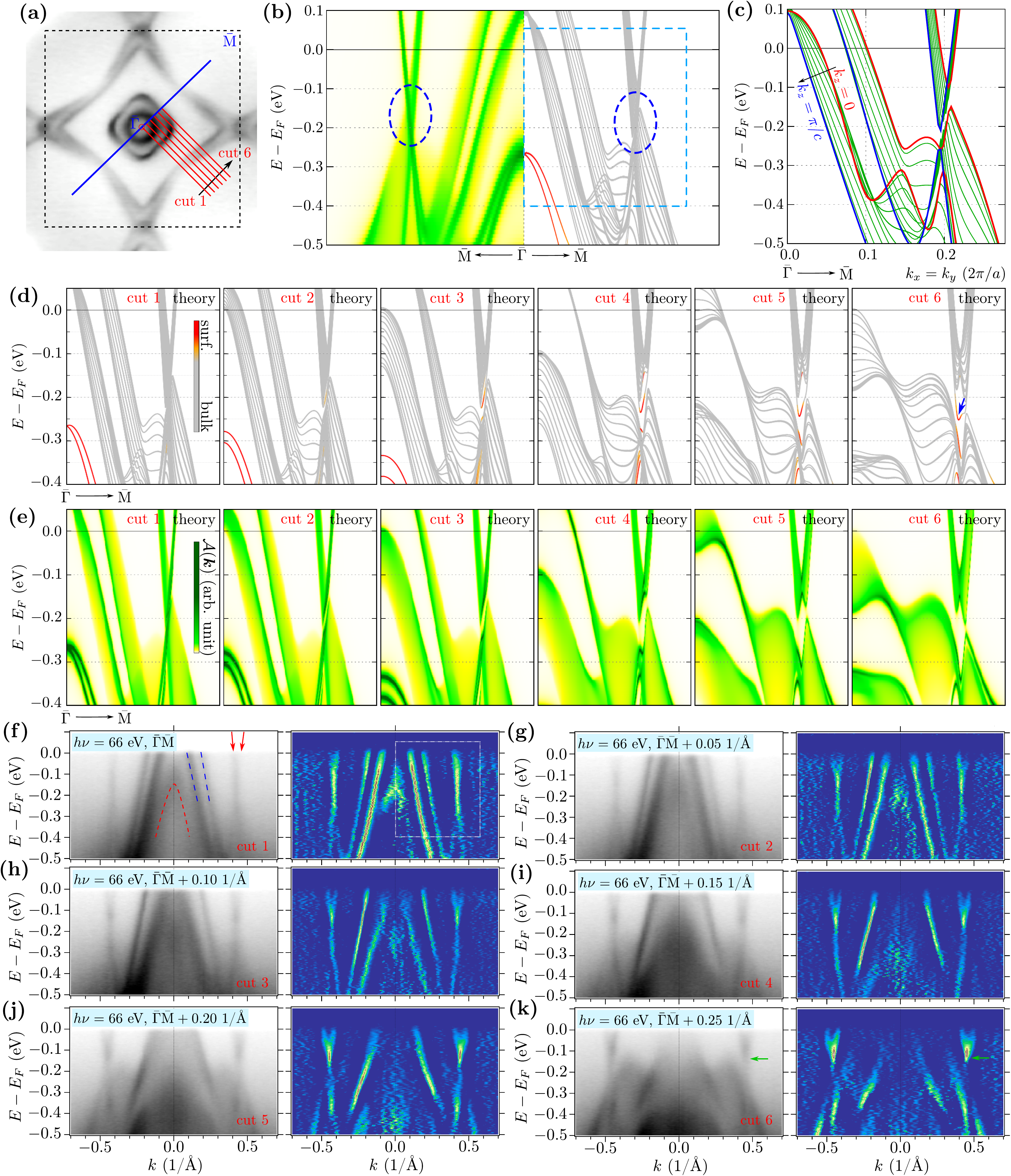}
\caption{The electronic band structure along a series of paths presented in (a).
Theoretical spectral function (left) and band structure (right) along the $\bar{\Gamma}$--$\bar{\text{M}}$ path for finite system (b) and related bulk band structure (c).
Panels (d) present band structures calculated for the slab geometry including both LaSb and Sb terminations, along different cuts shown in (a).
Bands with red (grey) color correspond to the surface (bulk) states.
Theoretical prediction of the spectral function within the surface Green's function calculation for LaSb termination, along different cuts from (a) is presented in (d).
Panels (d) and (e) correspond to the range of parameters marked by boxes in (b) and (f).
Panels (f)--(k) show the experimental spectra recorded with the ARPES measurements with $h\nu = 66$~eV at a temperature of $12$~K (left panels) and their 2D curvatures (right panels), along different cuts from (a). 
Blue dashed lines in (f) indicate the bands forming central pockets of the FSs, while red dashed line refers to the surface state. 
Red arrows in (f) indicate the position of the band corresponding to the diamond-like part of the FS and green arrows in (k) point the bottom of the band above the energy gap related to blue arrow on cut 6, i.e. panel (d).
}
\label{fig.arpes_cut_k_gm}
\end{figure*}

The band structure along the $\bar{\Gamma}$--$\bar{\text{X}}$ direction is presented in Fig.~\ref{fig.dnl1}.
As it can be seen, in the experimental spectra we can well separate visible bands.
The most intensive lines demonstrate the shape of the projected FS in a form of a pair of cylindrical pockets centered at $\bar{\Gamma}$ and a pair of diamond-like pockets [cf.~Fig.~\ref{fig.fbz}(a)].
Additionally, exactly at the $\Gamma$ point, around $-0.2$~eV we can see the surface states in a form of the parabolic like band [highlighted by a dashed white line in Fig.~\ref{fig.dnl1}(i)].
Observation of this surface state is in agreement with the direct DFT calculations (cf. discussion in Sec.~\ref{sec.slabdft}) and surface Green function calculations [surface state is marked by blue arrow in Fig.~\ref{fig.dnl1}(b)]

Additionally, examination of the band structure along the $\bar{\Gamma}$--$\bar{\text{X}}$ direction uncovers also other features of the band structure discussed earlier.
The NLs are visible as a crossing of the bands at the $\bar{\text{X}}$ point.
This crossing is visible independently of the energy of photons used during the ARPES experiment (red dashed lines on right panels in Fig.~\ref{fig.dnl1}). 
ARPES scans with variable photon energy yield constant energy intensity maps as a function of $k_{\perp} = k_{z}$ and $k_{\parallel} = k_{x}$ (perpendicular and parallel to the surface component of the wave vector) in $\Gamma$--X--R--Z plane and the data are compared to the calculated spectral function [see from Fig.~\ref{fig.dnl1}(c) to \ref{fig.dnl1}(f)]. 
The data are shown for $E_F$ [cf. Fig.~\ref{fig.dnl1}(c) and Fig.~\ref{fig.dnl1}(d)] and $-0.2$~eV below $E_F$ [cf. Fig.~\ref{fig.dnl1}(e) and Fig.~\ref{fig.dnl1}(f)]. 
The spectra have been collected for photon energies in the range from $20$~eV to $84$~eV and $k_{\perp}$ was obtained with the assumed inner potential of $V_{0}=15$~eV.
The comparison of the spectra and theory may indicate the actual assignment of high symmetry points as proposed in Fig.~\ref{fig.dnl1}(c)--\ref{fig.dnl1}(f). 
Taking into account this last result, as well as the excellent agreement of experimental and theoretical data (e.g. Fig.~\ref{fig.arpes_cut_x}), we conclude that the $h\nu = 66$~eV corresponds to the $k_{z}$ close to the $\Gamma$--X--M plane.
Indeed, the existence of the band crossing independently on $k_{\perp}$ confirms the realization of NL at $\bar{\text{X}}$ (what has been discussed earlier).

Moreover, around $- 0.1$~eV we can observe a blurred shape [marked by red arrow in Fig.~\ref{fig.dnl1}(i)].
The theoretical study clearly shows that the additional structure in the spectra can be explained by the projection of the bulk band on the surface Brillouin zone.
Indeed, the calculated {spectral} function confirms this explanation [cf. red arrow in Fig.~\ref{fig.dnl1}(b)].
This structure was also observed in the form of a cylindrical shape in the isoenergetic cuts [e.g., marked by a black arrow in Fig.~\ref{fig.arpes_cute}(d)].

\subsection{Surface states along $\bar{\Gamma}$-$\bar{\text{M}}$}
\label{sec.surf_state_gm_dir}

In the bulk band structure there exist several places where the SOC leads to opening of the gap between initially crossed bands.
Many of them are located along the $\Gamma$--M and Z--R paths with energies close to the Fermi level.
The previous study of Shi {\it et al.}~\cite{shi.richard.16} suggested the realization of the Dirac-cone-like structure in the vicinity of the Fermi level formed by the crossing of two linear energy bands.
In this section, we present a complex study of the electronic band structure, based on theoretical and experimental results, along different cuts parallel to $\Gamma$--M (Fig.~\ref{fig.arpes_cut_k_gm}).

The surface Green function and slab band structure are presented in Fig.~\ref{fig.arpes_cut_k_gm}(b) (left and right half, respectively).
The crossing of the linear bands around $-0.15$~eV is well visible in both results (marked by blue dashed ellipses).
Projection of the bulk electronic bands on the surface Brillouin zone, directly presents $k_{z}$ dependence of separate bands [Fig.~\ref{fig.arpes_cut_k_gm}(c)].
Independently on $k_{z}$ the direct gap is realized in the bulk (around $k_{x} = k_{y} \simeq 0.2 \times 2\pi/a$).
The value of this gap can be estimated as $\sim 5$~meV, which can be seen in high resolution ARPES experiment.
Additionally, this gap changes its value for directions away from the $\Gamma$--M path (cf. Fig.~\ref{fig.bulk_bands_gm} in SM~\cite{Note1}).
Moreover, the bulk gap is shifted from energy $-0.125$~eV to $ -0.225$~eV below the Fermi level.

Direct analysis of the band in slab geometry (including both LaSb and Sb terminations) is presented in Fig.~\ref{fig.arpes_cut_k_gm}(d), where the color of line denotes contribution of the surface or bulk states (red and grey color, respectively).
Similarly like previously the surface states at the $\bar{\Gamma}$ point are visible, which is realized in a system with the LaSb termination.
The other surface state (marked by the blue arrow) is realized by the second surface of the slab, terminated by the Sb square net.
Here we should have in mind that we have not observed this latter termination in our system.
Indeed, the surface Green function calculated for the LaSb termination [Fig.~\ref{fig.arpes_cut_k_gm}(e)], correctly reproduces the surface state found at the $\bar{\Gamma}$ point. At the same time, the other surface state is not observed.
The corresponding experimental results for different cuts are presented in panels from Fig.~\ref{fig.arpes_cut_k_gm}(f) to Fig.~\ref{fig.arpes_cut_k_gm}(k).
Firstly, we observe the disappearance of the surface state at the $\bar{\Gamma}$ point [red dashed line in Fig.~\ref{fig.arpes_cut_k_gm}(f)], while moving away from the $\Gamma$--M direction.
Secondly, we observe strong momentum-dependence of the bands forming central pockets of the FSs [two blue dashed lines in Fig.~\ref{fig.arpes_cut_k_gm}(f)].
For instance, at the cut 4, which crosses only the external $\bar{\Gamma}$-centered FS, we clearly see the shift of the internal branches to lower energies.
Two branches corresponding to diamond-like part of the FS [marked by two red arrows in Fig.~\ref{fig.arpes_cut_k_gm}(f)] also yield slightly different dispersions.
These two branches should play a role in realization of the discussed surface state.
Directly along the $\bar{\Gamma}$--M direction (cut 1) a cross-like structure is visible. Away from this cut, the open gap becomes more visible, where the bottom of the upper branches is visible [marked by a green arrow in ig.~\ref{fig.arpes_cut_k_gm}(k)].

The surface Green function correctly reproduces the experimental results.
In both cases, the surface state closing the gap is not visible.


\section{Summary}
\label{sec.summary}

In this paper, we have performed the systematic high resolution ARPES measurements on LaAgSb$_{2}$ covering a large area of the Brillouin zone.
The presented experimental results were supported by the theoretical analyses, within the {\it ab initio} (DFT) calculations, as well as by the tight binding model formulation.
Both theoretical techniques allow for excellent reproduction of the experimentally observed band structure of LaAgSb$_{2}$ (cf. Sec.~\ref{sec.bulkband}). 
Nodal lines related to the symmetry of the system are shown in calculations and corresponding ARPES spectra are obtained. 
Moreover, analyses of the band structure uncover the highly degenerated nodal lines, the existence of which is not a consequence of the {\it P4/nmm} symmetry.
These nodal lines occur due to vanishing the spin--orbit coupling, and they are located at the boundary of the Brillouin zone.

Direct studies of the system with slab geometry indicates a possibility of realization of the surface states in this compound (Sec.~\ref{sec.slabdft}).
We conclude that the appearance of the surface states strongly depends on the termination.
For instance, the surface states can be realized in the form of well visible separate parabolic-like bands, when the sample is not terminated by the Sb layer.
Indeed, our experimental data confirm
this theoretical prediction (Sec.~\ref{sec.surf_state_gx}).
We find the surface states at the $\Gamma$ and X points in the form of parabolic-like bands. 
However, this state disappears relatively fast due to adsorption of surface contamination.
Nevertheless, the observed surface states at the $\Gamma$ and X points indicate the LaSb termination of the studied sample.

\begin{figure}[!t]
\includegraphics[width=\linewidth]{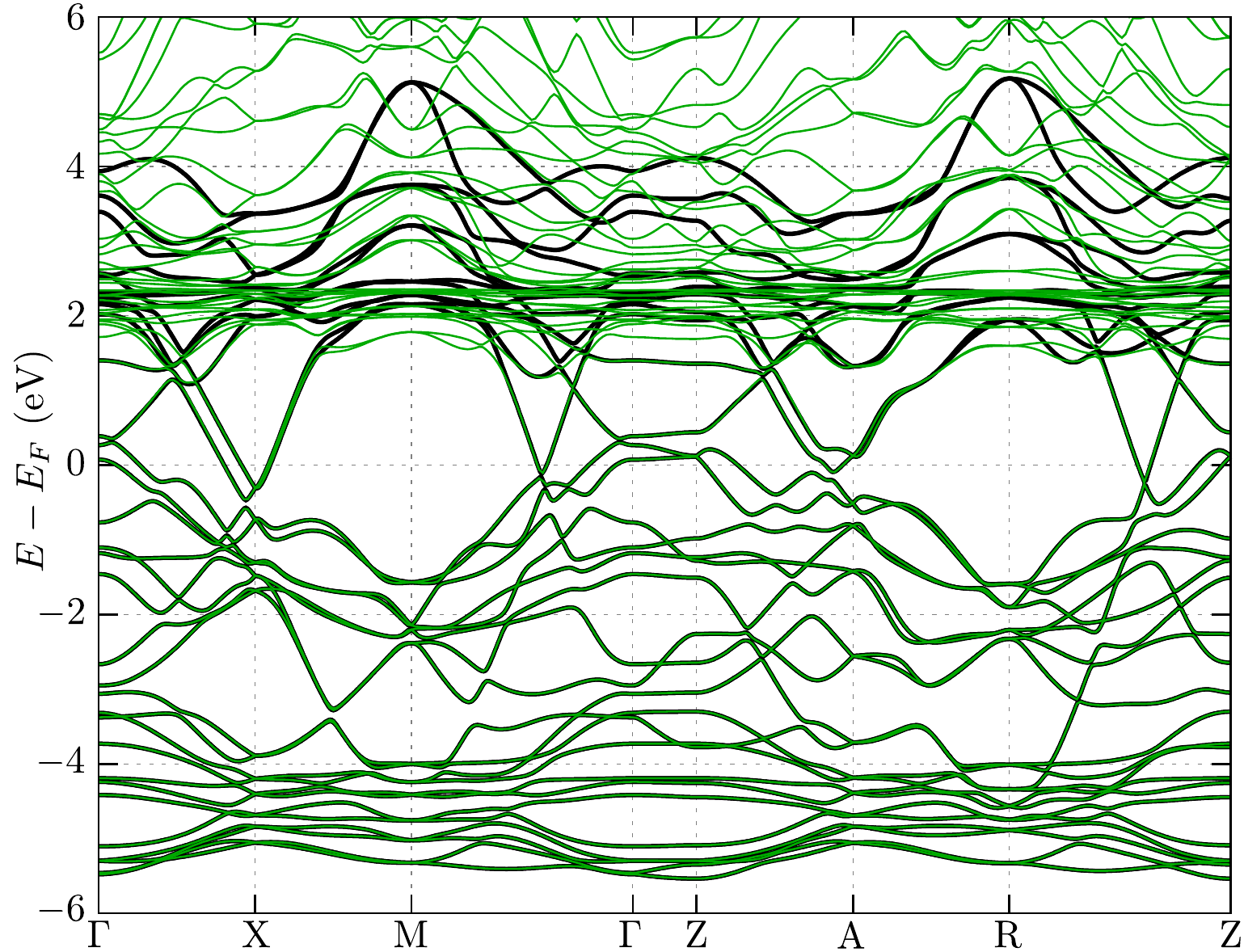}
\caption{
The comparison of the band structure obtained from the DFT calculations (green lines) and tight binding model (black lines).
}
\label{fig.band_compare}
\end{figure}

Finally, we 
verified if the Dirac surface states can be found along the $\Gamma$--M direction. 
Careful analysis of this problem was presented in Sec.~\ref{sec.surf_state_gm_dir}. 
The theoretical examination of the band structure clearly shows that the Dirac surface state can appear only at the Sb square net termination, in contrast to LaSb termination found in our samples. As a consequence the Dirac surface states along $\Gamma$--M are not observed.
Similar results can be obtained for the surface Green function analyses, where the surface state is also absent.

\begin{acknowledgments}
Some figures in this work were rendered using {\sc Vesta}~\cite{momma.izumi.11} and {\sc xCrysDen}~\cite{kokalj.99}.
Support of the Polish Ministry of Science and Higher Education under the grant N17/MNS/000039 is acknowledged.
This work was supported by the National Science Centre (NCN, Poland) under grants No. 
2017/25/B/ST3/02586 (P.P)
and
2017/24/C/ST3/00276 (A.P.).
A.P. appreciates funding in the frame of scholarships of the Minister of Science and Higher Education (Poland) for outstanding young scientists (2019 edition, no. 818/STYP/14/2019).
\end{acknowledgments}


\appendix

\begin{figure}[!t]
\includegraphics[width=\linewidth]{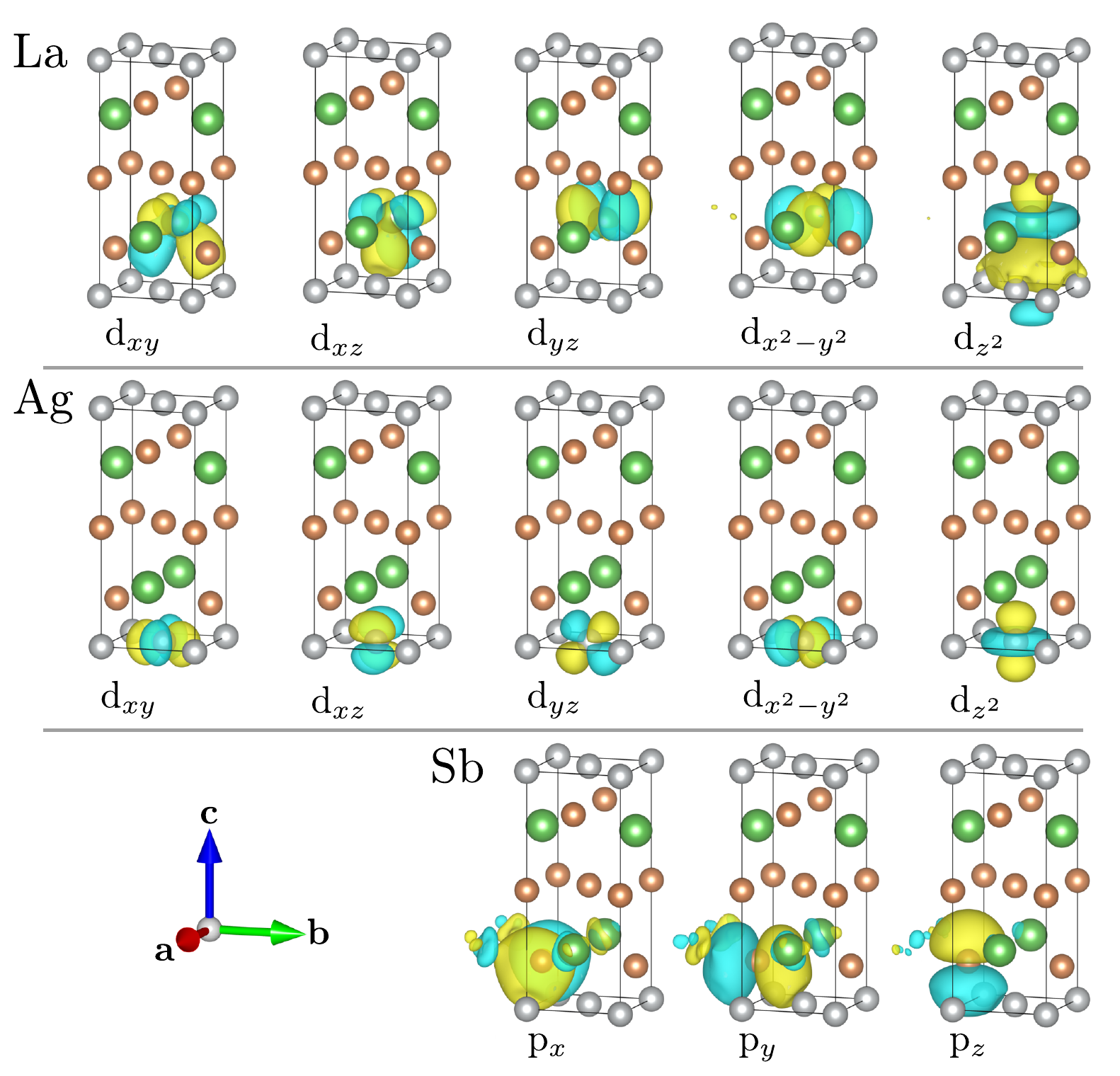}
\caption{
The final form of the maximally localized Wannier orbitals, found during procedure of the DFT band structure wannierization.
Groups from top to bottom correspond to $d$-like orbitals of Ag and La, and $p$-like orbitals of Sb, as labeled.
}
\label{fig.wan}
\end{figure}

\section{Tight binding model in maximally localized Wannier orbitals basis}
\label{sec.model_tbm}

Using results of the DFT calculation for electronic band structure we can find the tight binding model in the basis of the maximally localized Wannier orbitals~\cite{marzari.mostofi.12,marzari.vanderbilt.97,souza.marzari.01}. 
It can be performed via the {\sc Wannier90} software~\cite{mostofi.yates.08,mostofi.yates.14,pizzi.vitale.20}.
As a result of this we can find the parameters of the tight binding Hamiltonian of free electrons in the form:
\begin{eqnarray}
\mathcal{H} = \sum_{{\bm R},{\bm R}'} \sum_{\mu,\mu'} \sum_{\sigma,\sigma'} t_{{\bm R}\mu\sigma,{\bm R}'\mu'\sigma'} c_{{\bm R}\mu\sigma}^{\dagger} c_{{\bm R}'\mu'\sigma'},
\end{eqnarray}
where $c_{{\bm R}\mu\sigma}^{\dagger}$ ($c_{{\bm R}\mu\sigma}$) denotes the creation (annihilation) operator of the electron with spin $\sigma$ at $\mu$ orbital localized on atom in ${\bm R}$ position.
Here $t_{{\bm R}\mu\sigma,{\bm R}'\mu'\sigma'}$ for $\sigma = \sigma'$ ($\sigma \neq \sigma'$) denotes the spin conserved (spin flip) hopping integrals between orbitals $\mathcal{O}_{1} = ({\bm R}\mu)$ and $\mathcal{O}_{2} = ({\bm R}'\mu')$.
We should notice that the spin flip hopping part corresponds to the SOC, i.e. mixing of the different type of spin subspaces. 
From this, the normal states in momentum space can be rewritten in the matrix form:
\begin{eqnarray}
\label{eg.ham_matrix}
\mathbb{H}_{\mu\sigma,\mu'\sigma'} \left( {\bm k} \right) = \sum_{\bm d} \exp \left( i {\bm k} \cdot {\bm d} \right) t_{{\bm R}\mu\sigma,{\bm R}'\mu'\sigma'} ,
\end{eqnarray}
where ${\bm d} = {\bm R} - {\bm R}'$ is the real-space distance between orbital $\mathcal{O}_{1}$ and $\mathcal{O}_{2}$.
The band structure for a given ${\bm k}$-point can be found by diagonalization of the matrix $\mathbb{H} ( {\bm k} )$.

In our case, the tight binding model was found from the $8 \times 8 \times 6$ full ${\bm k}$-point DFT calculation, starting from $p$ orbitals of Sb, as well as $d$ orbitals of Ag and La. 
This gives us $32$-orbital tight binding model of the LaAgSb$_{2}$.
Comparison of the band structure obtained from the DFT and tight binding model can be found in Fig.~\ref{fig.band_compare}, while the Wannier (maximally localized) orbitals are presented in Fig.~\ref{fig.wan}.

\newpage


\bibliography{biblio.bib}


\clearpage
\newpage

\onecolumngrid


\begin{center}
\textbf{\large Supplemental Material:\\[.2cm] Electronic band structure and surface states in Dirac semimetal LaAgSb$_{2}$}\\[.5cm]
Marcin Rosmus,$^{1,2}$ Natalia Olszowska,$^{2}$ Zbigniew Bukowski$^{3}$, Pawe\l{} Starowicz$^{1}$, Przemys\l{}aw Piekarz$^{4}$, Andrzej Ptok$^{4}$ \\[.3cm]
{\itshape
\mbox{$^{1}$Marian Smoluchowski Institute of Physics, Jagiellonian University, Prof. S. {\L}ojasiewicza 11, PL-30348 Krak\'{o}w, Poland}\\[.1cm]
\mbox{$^{2}$Solaris National Synchrotron Radiation Centre, Jagiellonian University, Czerwone Maki 98, 30-392 Krak\'{o}w, Poland}\\[.1cm]
\mbox{$^{3}$Insitute of Low Temperature and Structure Research, Polish Academy of Sciences, P.O. Box 1410,50-950 Wroc\l{}aw, Poland}\\[.1cm]
\mbox{$^{4}$Institute of Nuclear Physics, Polish Academy of Sciences,
ul. W. E. Radzikowskiego 152, PL-31342 Krak\'{o}w, Poland}
}\\[.1cm]
(Dated: \today)
\\[1cm]
\end{center}

\setcounter{equation}{0}
\renewcommand{\theequation}{S\arabic{equation}}
\setcounter{figure}{0}
\renewcommand{\thefigure}{S\arabic{figure}}
\setcounter{section}{0}
\renewcommand{\thesection}{S\arabic{section}}
\setcounter{table}{0}
\setcounter{page}{1}

\begin{figure}[!h]
\includegraphics[width=\linewidth]{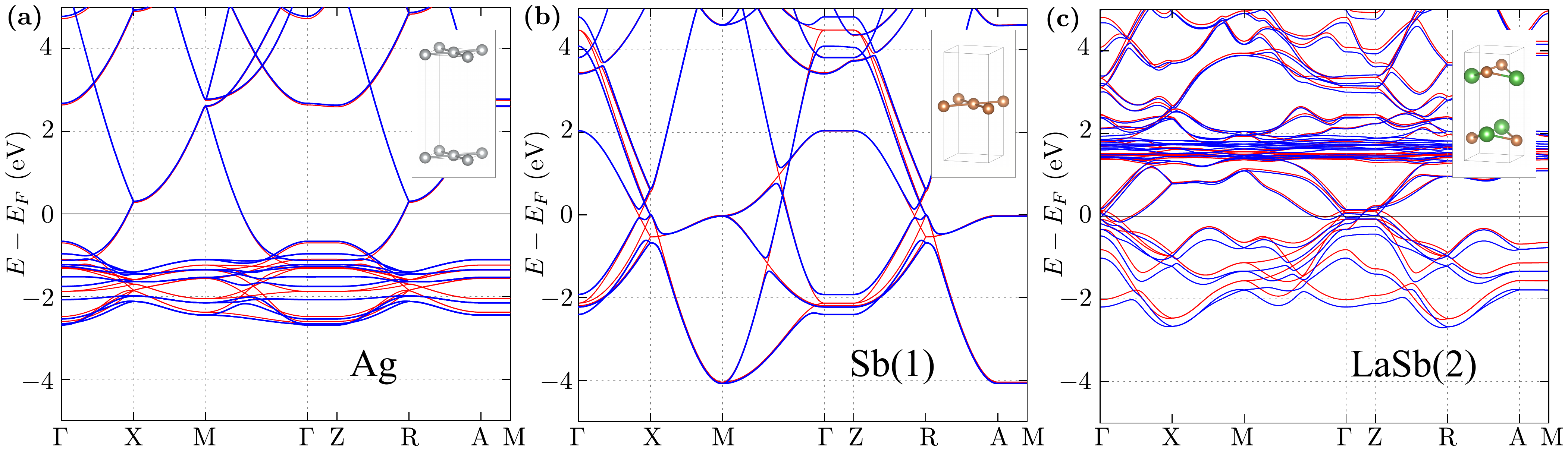}
\caption{
The bulk electronic band structure of layers forming bulk LaAgSb$_{2}$ structure (as labeled and shown in insets).
Bands crossing the Fermi level in the case of Ag and Sb nets, correspond mostly to the $p$ orbitals -- which lead to the characteristic diamond--like Fermi surface with corner at X points.
The Fermi surface centered at $\Gamma$ is associated with bands of 
LaSb$_{2}$ double-layer showing nonsymmorphic glide symmetry.
}
\label{fig.bands_layer}
\end{figure}

\begin{figure}[!b]
\includegraphics[width=0.75\linewidth]{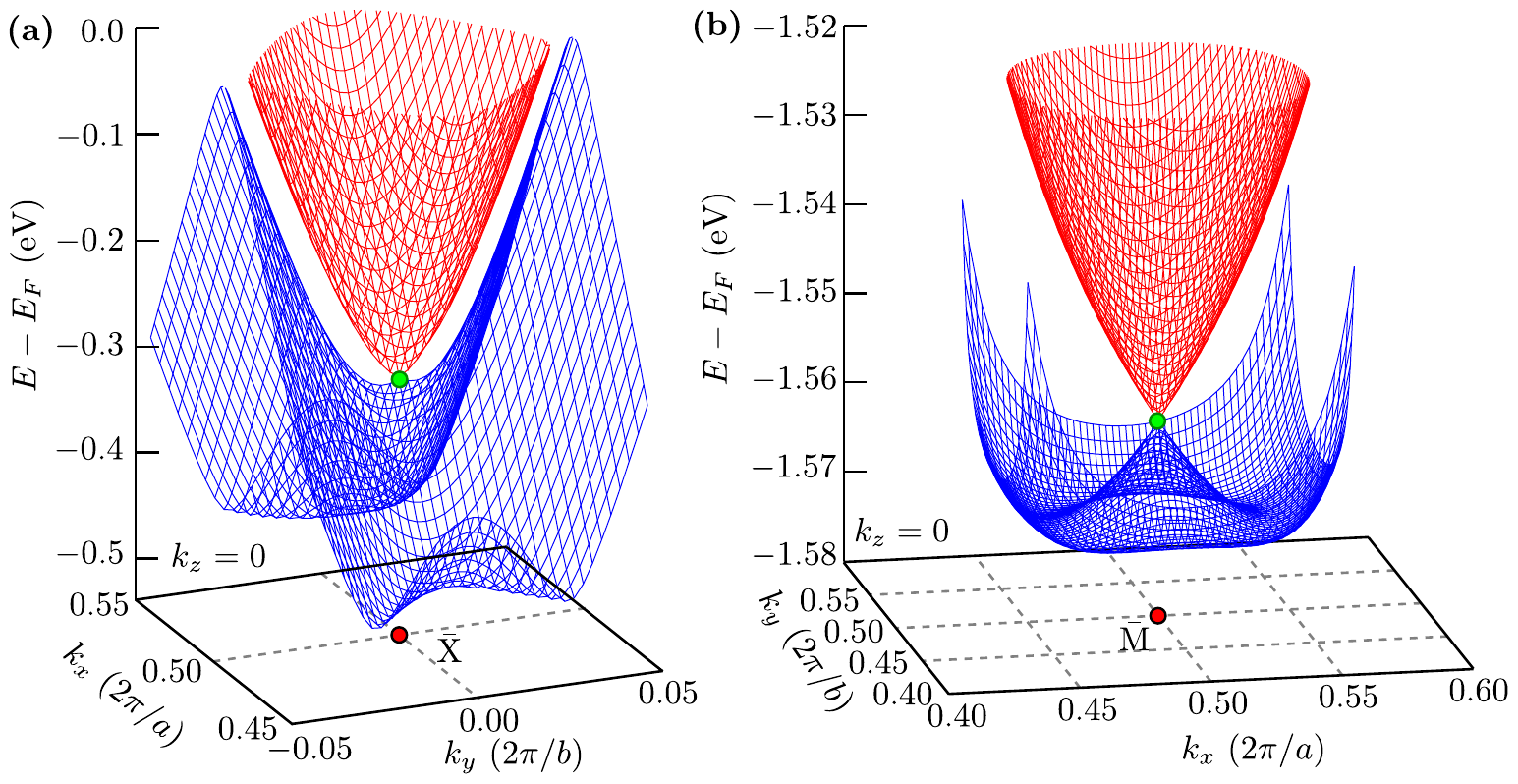}
\caption{ 
A realization of a Dirac point (green dot) at X (a) and M (b) points (zoom on insets at Fig.~\ref{fig.bulk_bands}) forming a nodal line along X--A and M--R directions, respectively.
In the case of the X point (a), the Dirac point is formed between the bottom vertex of the upper band (red) and the saddle point of the lower band (blue).
In the case of M point (b), the Dirac point is realized between two bands a spliting of which is given by the spin--orbit coupling in the typical Rashba-like form. 
}
\label{fig.bulk_dirac}
\end{figure}

\begin{figure}[!t]
\includegraphics[width=\linewidth]{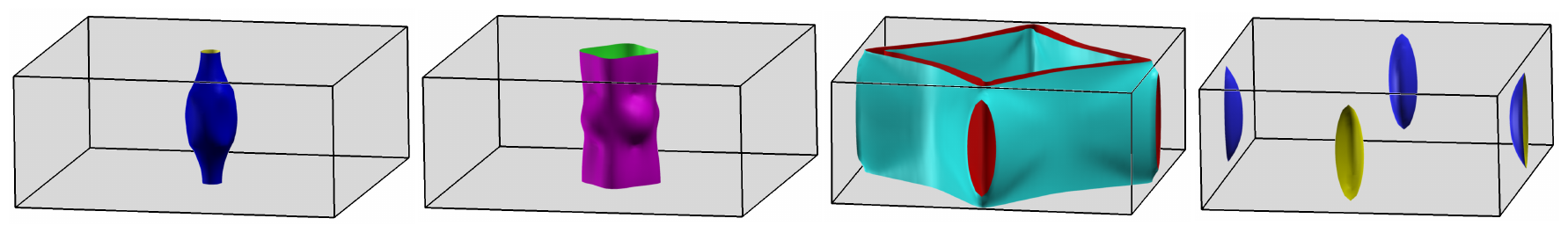}
\caption{
Branches of the Fermi surface of LaAgSb$_{2}$ shown in separate figures.
}
\label{fig.fs_com}
\end{figure}

\begin{figure}[!t]
\includegraphics[width=\linewidth]{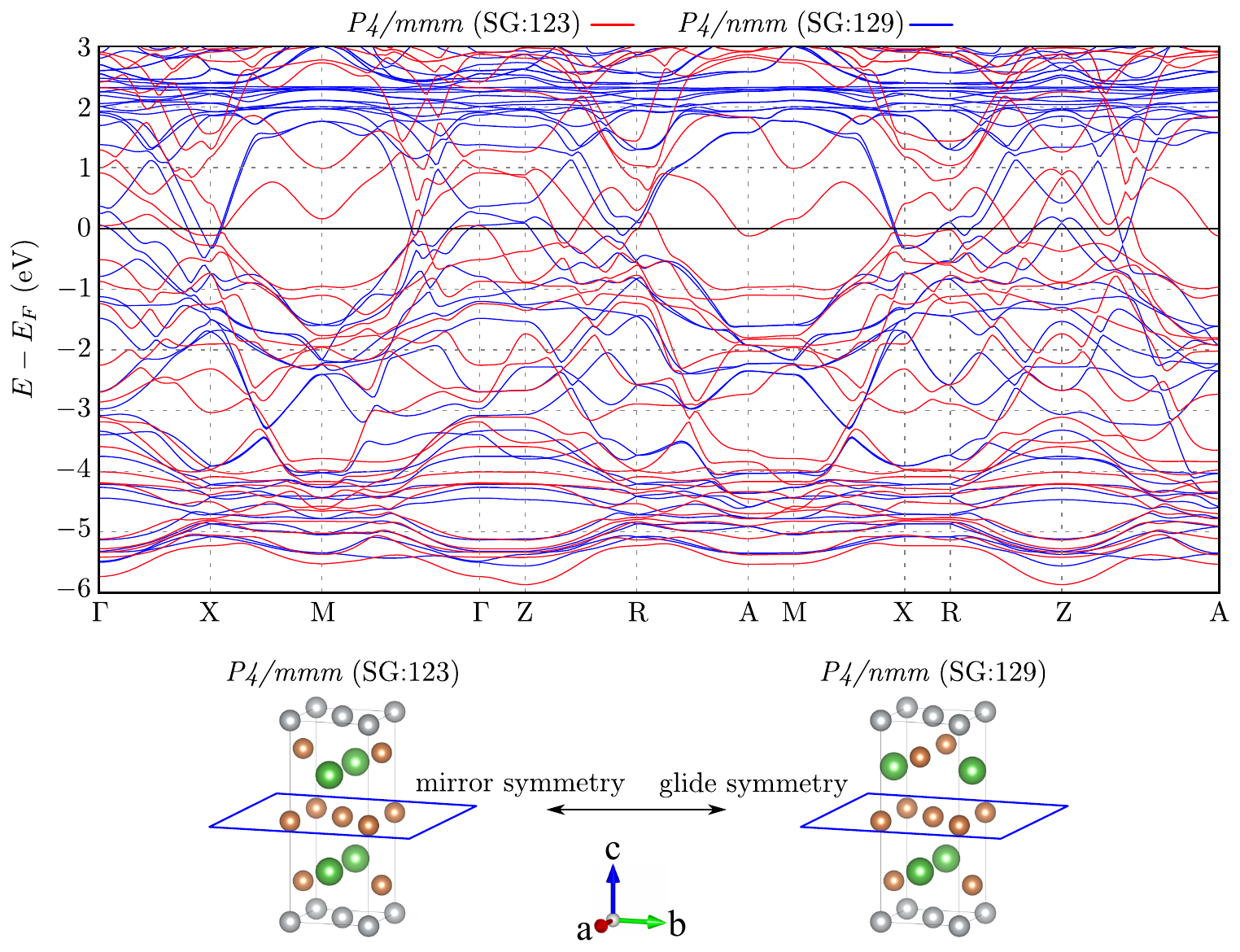}
\caption{
Comparison of the bulk electronic band structure of LaAgSb$_{2}$ for different structural variants, obtained from DFT calculations in the presence of the spin--orbit coupling.
Red and blue lines correspond to structure with {\it P4/mmm} and {\it P4/nmm} space group, respectively.
Replacement of the glide symmetry with the mirror symmetry (blue frame) causes a strong modification of the band structure of the system.
In practice, this leads to a disappearance of the Dirac cones at high symmetry points X, M, R, or A, as well as a removal of degeneracy along A--M and X--R paths.
}
\label{fig.bulk_bands_sg}
\end{figure}

\begin{figure}[!t]
\includegraphics[width=\linewidth]{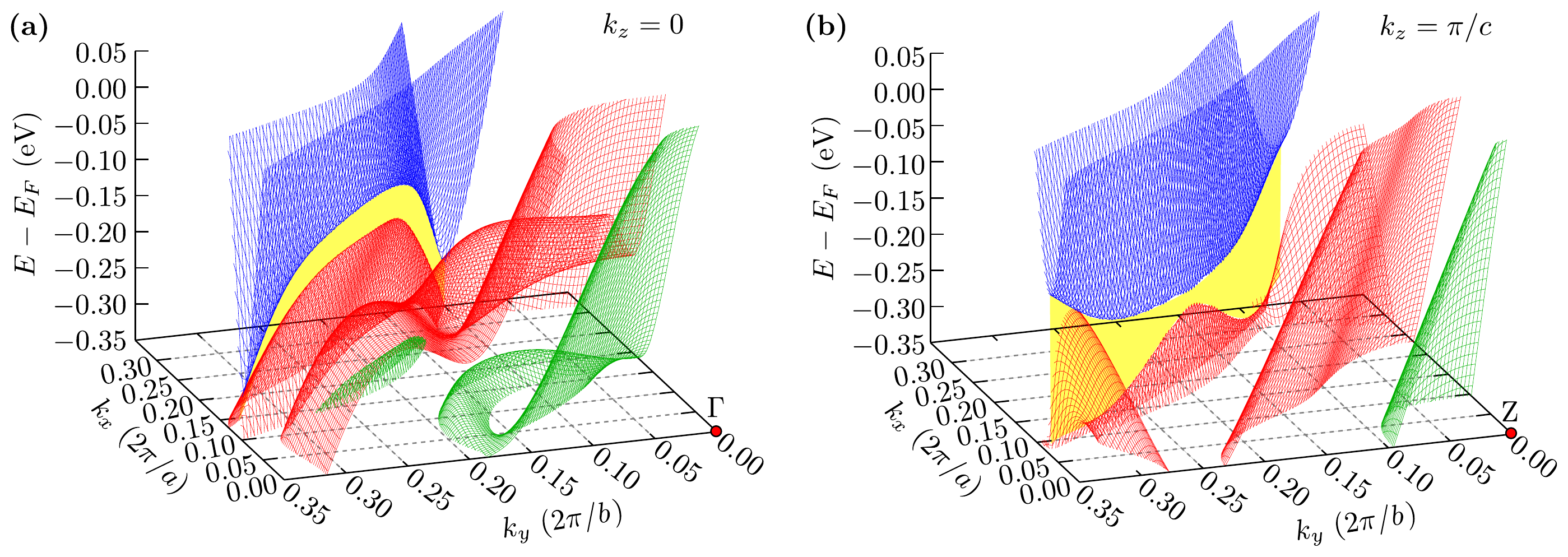}
\caption{
The bulk electronic band structure in the presence of the spin--orbit coupling along $\Gamma$--M path, for $k_{z} = 0$ (a) and $\pi/c$ (b).
The blue, red, and green surfaces correspond to the bands, while the yellow area shows the gap induced by spin--orbit coupling.
}
\label{fig.bulk_bands_gm}
\end{figure}

In this Supplemental Material we present additional band structures obtained from DFT calculations, in particular:
\begin{itemize}
\item Fig.~\ref{fig.bands_layer} -- the bulk electronic band structure related to layers forming LaAgSb$_{2}$ structure.
\item Fig.~\ref{fig.fs_com} -- Fermi surface branches shown separately.
\item Fig.~\ref{fig.bulk_bands_sg} -- comparison of the band structures in {\it P4/mmm} ({\it not realized in reality}, SG:123) and {\it P4/nmm} (SG:129) showing impact of the mirror symmetry and glide symmetry.
\item Fig.~\ref{fig.bulk_dirac} -- the Dirac points at X and M.
\item Fig.~\ref{fig.bulk_bands_gm} --the bulk electronic band structure along $\Gamma$--M path for $k_{z} = 0$ and $k_{z} = \pi/c$.
\end{itemize}

\end{document}